\documentclass[12pt]{article}
\def\la{\mathrel{\mathpalette\fun <}}

\def\fun#1#2{\lower3.6pt\vbox{\baselineskip0pt\lineskip.9pt
\ialign{$\mathsurround=0pt#1\hfil##\hfil$\crcr#2\crcr\sim\crcr}}}

\title{$\mbox{\boldmath$\mu_\nu$}$ \\
Review talk at the workshop ``Search for Dark Matter and Neutrino
Magnetic Moment'', \\ ITEP, 11.12.2001}
\author{M. Vysotsky \\
ITEP, Moscow 117218}
\date{}
\begin{document}
\large
 \maketitle

 \begin{abstract}

The present situation and hopes on bounding (founding) neutrino
magnetic moment in future are reviewed.

\end{abstract}

\section{Motivation: Solar neutrinos}

The reviewed story starts from the solar neutrino deficit and
apparent anticorrelation of the Homestake data on the solar
$\nu_e$ flux with the Sun magnetic activity \cite{1}. To explain
this anticorrelation the hypothesis of a large neutrino magnetic
moment was suggested in \cite{2}. According to it the left-handed
electron neutrinos, produced in thermonuclear reaction in the Sun
core, are (partly) transformed to the right-handed neutrinos when
they pass the toroidal magnetic field generated in a solar
convective zone in the years of active Sun. (It was noted in
\cite{3} that the electric dipole moment of ultrarelativistic
neutrino would lead to the same effect.) This field manifests
itself as the Sun spots -- low temperature regions on the Sun
surface where a toroidal field goes out from (or comes inside) the
Sun. The number of the left-handed neutrinos which survive is
given by the following formula:
\begin{equation}
N_L = N_0 \cos^2[\mu_\nu \int H_\bot dx] \;\; , \label{1}
\end{equation}
where $N_0$ is the initial flux, $H_\bot$ is the component of the
magnetic field normal to the neutrino trajectory and the integral
goes along a straightforward neutrino trajectory. For the toroidal
field $H_\bot \gg H_\parallel$, that is why formula (\ref{1}) is
valid. It is convenient to measure $\mu_\nu$ in Bohr magnetons,
$\mu_B = e/2m_e \approx 3 \cdot 10^{-4} \frac{1}{{\rm G} \cdot
{\rm cm}}$. The width of the solar convective zone $L$ equals
approximately $2 \cdot 10^{10}$ cm. The magnitude of the toroidal
magnetic field is not known. At the Sun spots it varies between 2
and 4 kG and in some solar models the toroidal field grows at
inner regions of the Sun. There exists the upper bound: the
magnetic field inside the Sun can not exceed $\sim 100$ kGauss.
That is why in order to have considerable reduction of the active
electron neutrino flux $\mu_\nu$ should be bounded from below by
at least
\begin{equation}
\mu_\nu > 10^{-12} \mu_B \label{2}
\end{equation}
The toroidal field depends on the time with the 22 years period
reaching the maximum values each 11 years, at the periods of the
active Sun.

At the years of the quiet Sun the toroidal field transforms to
poloidal; the field configuration is that of a dipole. The
magnitude of the poloidal field is several orders less than that
of a toroidal field. The directions of the toroidal field are
opposite at the northern and southern solar hemispheres. That is
why in the vicinity of a solar equator the toroidal field vanishes
even when the Sun is active. Due to the inclination of the Sun's
rotation axis to the ecliptic we come to the prediction of a
half-a-year period of the electron neutrino flux in the years of
the active Sun \cite{4}. The traces of this periodicity were found
in Homestake data \cite{5}. Finally, in paper \cite{6} the damping
of the spin flip due to a neutrino interaction with the matter was
considered.  We note also in that paper that the existence of a
sterile right-handed neutrino is not necessary for the phenomenon
to occur because the muon (or tau) antineutrinos could play its
role  in  case of the so-called Majorana magnetic moment (see also
\cite{66}).

Our papers were not the first where the influence of the solar
magnetic field on the flux of neutrinos from the Sun was analyzed.
In paper \cite{7} it was found that for the solar magnetic field
of the order of $10^6$ Gauss the flux of active neutrinos would be
reduced for $\mu_\nu > 10^{-13} \mu_B$. However, the time
variation of a solar neutrino flux was not considered in \cite{7}.

The azimuthal angle distribution of electrons on which the solar
neutrinos scatter could help to reveal the neutrino spin rotation
inside the Sun \cite{88}.

\section{Bounds on $\mbox{\boldmath$\mu_{\nu_e}$}$}

They arrive from the astrophysical  considerations and experiments
with reactor $\bar\nu_e$. The first are more stringent while the
second ones -- more reliable.

The most restrictable bound has come from the consideration of a
supernova explosion. Trapped in a supernova interior, the active
neutrinos diffuse to the star shell approximately 10 seconds and
this time interval coincides with the duration of a neutrino
signal observed at the moment of SN 1987A explosion by Kamiokanda
and IMB detectors. Trapping  occurs due to the weak interactions
of neutrinos. If neutrino has a nonzero magnetic moment, then the
scattering due to the photon exchange between a neutrino and a
charged particle in plasma leads to the neutrino spin flip. If a
produced particle is a right-handed neutrino sterile in a weak
interaction, it leaves SN without further interactions. This
pattern contradicts to the observed neutrino signal of SN 1987A.
The energy released in SN implosion is taken away by sterile
neutrinos. Due to this no energy is left for the envelope
explosion. To avoid these difficulties according to \cite{8}, the
neutrino magnetic moment should be bounded from above:
\begin{equation}
\mu_\nu^{SN} \la \sim 10^{-12} \mu_B \;\; . \label{3}
\end{equation}

The simplest way to avoid this bound is to use Majorana neutrino
magnetic moment which transfers a left-handed neutrino to an
anti-left-handed neutrino of another flavour. Both participate in
weak interactions and are trapped in supernova. Esthetically this
kind of a magnetic moment is much more appealing: we avoid
introduction of right-handed neutrinos, needed only for solution
of one particular problem. This scenario works with solar
neutrinos if the mass difference of two states mixed by the
magnetic field is bound from above \cite{6}:
\begin{equation}
\frac{\Delta m^2}{2E} \la \mu H \;\; . \label{4}
\end{equation}

For $\mu = 10^{-11} \mu_B$, $H = 10^4$ G and $E = 10$ MeV we
obtain $\Delta m^2 \la 10^{-8}$ eV$^2$.

However, even the case of Dirac neutrino could work: the magnetic
field inside supernova can flip neutrino spin back transforming
them into active left-handed particles. This mechanism could help
to transfer the energy to star envelope solving the problem of
supernova explosion \cite{9}.

The next set of bounds comes from an additional star cooling
mechanism due to a plasmon decay into a neutrino pair. Such a
mechanism would essentially change the time evolution of stars.
This contradicts the observed temperature dependence of the star
population unless $\mu_\nu$ is small enough \cite{10}:
\begin{equation}
\mu_\nu^{\rm star ~ cooling} \la \sim 10^{-11} \mu_B \label{5}
\end{equation}

This bound follows from the analysis of white dwarfs, red giants
and helium burning stars. Unlike the case of supernova, the
neutrinos leave these stars without scattering, so the only way to
avoid bound (\ref{5}) is to make the neutrino mass larger than the
plasma frequency, in this way making neutrino production in
plasmon decay kinematically forbidden. Since the plasma frequency
is of the order of keV, the electron neutrinos produced in the Sun
are definitely lighter and bound (\ref{5}) applies to the
phenomena in which solar neutrinos participate.

Finally, we come to the bounds from the experiments with the
reactor neutrinos. If they have a nonzero magnetic moment, then in
addition to scattering due to $W$- and $Z$-boson exchanges, one
should take into account the photon exchange. The weak and
electromagnetic scatterings do not interfere as far as neutrino
mass can be neglected. For the differential cross section of the
electron antineutrino scattering on the electron we obtain: $$
 \frac{d\sigma}{d T}  =
\left(\frac{\mu_\nu}{\mu_B}\right)^2 \frac{\pi\alpha^2}{m_e^2}
\left(\frac{1}{T} -\frac{1}{E_\nu}\right) +\frac{2G_F^2 m_e}{\pi}
\left[(1-\frac{T}{E_\nu})^2 g_L^2 + \right. $$
\begin{equation}
+ \left. g_R^2 - g_L g_R \frac{m_e T}{E_\nu ^2}\right] \; , \;\;
g_R = s_W^2 \; , \;\; g_L = \frac{1}{2} + s_W^2 \;\; , \label{6}
\end{equation}
where the first term is due to the photon exchange, the second one
is due to the weak interactions; $E_\nu$ is the energy of the
initial neutrino, $T$ -- the kinetic energy of recoil electron,
$G_F \approx 10^{-5}/m_p^2$ -- the Fermi constant, $s_W^2 \approx
0.23$ -- the electroweak mixing angle. We see that the relative
contribution of the photon exchange grows with diminishing of
neutrino energy, that is why one should look for the sources of
soft neutrinos in order to bound $\mu_\nu$ most effectively. Two
dedicated reactor experiments lead to the following bounds: $$
\mu_\nu^{\rm reactor} < 2.4 \cdot 10^{-10} \mu_B \;\; \mbox{\rm at
90\%  C.L.  Krasnoyarsk \cite{11}} $$
\begin{equation}
\mu_\nu^{\rm reactor} < 1.9 \cdot 10^{-10} \mu_B \;\; \mbox{\rm at
 95\%  C.L. Rovno \cite{12}} \;\; , \label{7}
\end{equation}
for review see \cite{144}.

Results of MUNU collaboration were recently announced. 60\% of
data are analyzed leading to the following bound: $$ \mu_\nu^{\rm
reactor} < 1.3 \cdot 10^{-10} \mu_B \;\; \mbox{\rm at 90\%  C.L.
Grenoble \cite{145}} \;\; . $$

One more reactor experiment with low threshold germanium detector
is running now \cite{146}.

In conclusion, comparing (\ref{2}) and (\ref{7}), we see that to
clarify if spin flip occurs when neutrinos cross the solar
magnetic field, the earth-based experiment which is sensitive to
the value of $\mu_\nu$ being two orders of magnitude smaller than
the existing reactor bound is highly desirable. The artificial
sources of low-energy neutrinos could provide radical progress in
this direction. Great expectations are connected with the
experiment with a tritium source since the neutrino energies are
very low, $E_\nu < 18$ keV, and a powerful source might be
available \cite{13}.

\section{Models}

The neutrino magnetic moments are zero in Standard Model with
three flavors of massless left-handed neutrinos. Dirac moments are
zero since there are no right-handed components while Majorana
moments are zero because the lepton quantum numbers are conserved.
If neutrino is a massive Dirac particle, then loop diagrams with
$W$ exchange lead to nonzero $\mu_\nu$:
\begin{equation}
\mu_\nu = \frac{eg^2}{16\pi^2} \frac{m_\nu}{M_W^2} \frac{3}{4}
\approx 3 \cdot 10^{-10} \frac{m_\nu}{m_p}\mu_B \;\; , \label{8}
\end{equation}
where $g$ is $SU(2)_L$ gauge coupling constant and $m_p$ -- the
proton mass. From the investigation of a tritium beta spectrum, we
know that $m_{\nu_e} < 1$ eV, that is why $\mu_\nu$ described by
eq. (\ref{8}) is many orders of magnitude smaller than the one
interesting for the solar neutrino. The proportionality of
$\mu_\nu$ to the neutrino mass originates from left-handedness of
weak interactions. $W$-boson interacts only with the left-handed
fermions, that is why spin flip should occur on a neutrino line.

In the left-right symmetric extensions of $SU(2)_L \times U(1)$
theory or in a model with charged scalar interacting with leptons,
spin-flip can occur on a charged fermion line and substituting
mass of charged lepton ($e$, $\mu$ or $\tau$) instead of $m_\nu$
in eq. (\ref{8}) we get the value of $\mu_\nu$ which can lead to
spin flip of neutrinos in the Sun \cite{14}. However, all such
models have naturality problem. The point is that the same loop
diagrams which generate a neutrino magnetic moment  contribute to
neutrino mass when an external photon line is eliminated. This
contribution is logarithmically divergent. However, the
coefficient in front of the logarithm is:
\begin{equation}
\Delta m_\nu \sim \frac{\mu_\nu}{e}M^2 \approx 10^{-11}
\frac{M}{m_e} M \sim 100 \; {\rm keV} \;\; , \label{9}
\end{equation}
and at least five orders of magnitude should be fine tuned in
order to get $m_\nu < 1$ eV (we substituted $\mu_\nu = 10^{-11}
\mu_B$ and the mass of a heavy charged particle $M \approx 100$
GeV which is a lower bound from the absence of this particle in
LEP II experiments).

In paper \cite{15} it was pointed out that the naturality problem
could be avoided by an $SU(2)$ symmetry that would forbid neutrino
masses but allow the nonzero magnetic moment. The beautiful
realization of this idea was suggested in \cite{16}, where
horizontal $SU(2)_H$ symmetry between leptons of the first two
generations was used. The operator of Majorana magnetic moment,
being antisymmetric with respect to the permutations of $\nu_e^L$
and $\nu_\mu^L$, is $SU(2)_H$ singlet, while Majorana mass terms
are components of $SU(2)_H$ triplet. That is why $SU(2)_H$ forbids
Majorana neutrino masses while the magnetic moment is allowed. In
\cite{16} it is generated by the diagrams with a heavy charged
scalar propagated in the loop. $SU(2)_H$ is not an exact symmetry
since it is violated by the difference of the electron and muon
masses. This difference is tiny; multiplying the right-hand side
of equation (\ref{9}) by $(m_\mu -m_e)/M_W$, we avoid the
naturality problem.

The $SU(2)$ custodial symmetry which helps to avoid generation of
a too big neutrino mass in the models with large $\mu_\nu$ was not
the last word in model building. The generation of the magnetic
moment at a two-loop level was suggested in \cite{17}. One loop
generates $\gamma W S^+$ vertex, where $S^+$ is a charged scalar
particle. $S^+$ and $W$ are absorbed at a fermion line, leading to
Majorana neutrino magnetic moment. Removing a photon, we get a
two-loop contribution to the neutrino mass. The $S^+ W$ vertex
should be proportional to momentum $k_\mu$ which, acting on $W
e\nu$ vertex, converts into the charged lepton mass. We obtain
that the right-hand side of an estimate (\ref{9}) should be
multiplied by the factor $(m_e/M_W)^2$ removing in this way the
naturality problem (proportionality to the second power of lepton
mass follows from the well-known fact that changing a sign of
fermion mass we should get the same expressions for the
observables).

Concluding this part, we should say that a large number of
extensions of the Standard Model were suggested which lead to the
value of the neutrino magnetic moment, interesting from the point
of view of neutrino propagation  in the Sun. As a rule, these
models contain heavy charged scalars which interact with leptons.

\section{Solar neutrinos: sixteen years later}

A considerable progress in detecting solar neutrinos was achieved
after 1986. Together with Homestake experiment Kamiokanda, SAGE,
GALLEX, Super\-ka\-mio\-kan\-da, GNO and, finally, SNO experiments
were and are running. All of them, measuring the solar neutrino
fluxes in the different energy intervals, detect the neutrino
deficit in comparison to the standard solar model predictions. The
routine expla\-nation of this deficit has become neutrino
oscillation  considered many years ago by B. Pontecorvo \cite{18}.
In this section we shall analyze if the hypothesis of the neutrino
magnetic moment remains an alternative explanation of the neutrino
deficit. To do this let us look at the relevant papers, which
appeared during the last 16 years.

In paper \cite{19} the Homestake data obtained during the years
1970 -- 1991 were analyzed. This period covers two eleven-years
cycles of the solar activity. The authors studied the
anticorrelation of a neutrino flux with the solar surface magnetic
field. According to \cite{19}, the effect is very strong when the
magnetic field is taken in the vicinity of a solar equator, where
the neutrinos, which are detected on the Earth, pass. It
diminishes when the field at higher latitudes is taken into
account. Also Kamiokanda data from the period 1987 -- 1990 were
analyzed. At this period the Sun magnetic activity was rising
while no change in the neutrino flux was found in \cite{20}. The
authors of \cite{19} noted that there is approximately one year
delay in growing of the magnetic field at the low Sun latitudes
which could explain Kamiokanda result. Also we should note that
Majorana magnetic moment transforms electron neutrinos to muon or
tau antineutrinos which are sterile for Davies experiment (as well
as SAGE and GALLEX (GNO)) but scatter on electrons due to $Z$
exchange being active at Kamiokanda detector.

It was noted in paper \cite{21} that the solar magnetic field is
highly inhomogeneous and that some components of this field last
for several or even many Sun rotations. In view of this, the
authors look for periodicity in GALLEX-GNO data. The discovered
periodicity coincides with the rotation frequency of an equatorial
part of a convective zone confirming in this way the hypothesis of
neutrino flip by the magnetic field.

In a number of recent papers \cite{22} the data from all solar
neutrino detectors were analyzed in the framework of the neutrino
magnetic moment hypothesis. The general conclusion is that the fit
of the data of the same quality as that of using the neutrino
oscillations can be achieved. In the framework of $\mu_\nu$
scenario the energy dependence of neutrino suppression is achieved
when $\Delta m_\nu^2/E_\nu$ term in the neutrino Hamiltonian is
taken into account. Since the solar magnetic field varies along
the neutrino trajectory, resonance spin flip could occur (the
so-called Resonance Spin Flavor Precession \cite{23}; the neutrino
flavor is changed simultaneously with spin in case of Majorana
magnetic moments).

There is a general consensus in the literature that in order to
have observable effects on solar neutrinos the magnetic moment of
$\nu_e$ should be larger than $10^{-12} \mu_B$. In view of this
the laboratory experiment sensitive to $\mu_{\nu_e} \sim (10^{-11}
\div 10^{-12}) \mu_B$ is very actual. Projected experiment with a
powerful tritium source has been one of the topics of the present
workshop.

I am grateful to L.N. Bogdanova for the organization of this
workshop and for the invitation to present a review talk at it.

\end{document}